# Voltage Matching, Étendue, and Ratchet Steps in Advanced-Concept Solar Cells

Andreas Pusch [*] and Nicholas J. Ekins-Daukes
*School of Photovoltaic and Renewable Energy Engineering, UNSW Sydney, Kensington, 2052, Australia*

Many advanced solar cell concepts propose surpassing the Shockley-Queisser limit by introducing multiple quasi-Fermi-level separations that are arranged in series and/or in parallel. Exceeding the Shockley-Queisser limit with any parallel arrangement involves intermediate states that deliver additional charge carriers at, ideally, the same electrochemical potential as the other elements in the parallel network. This can be thought of as voltage matching individual parallel components, and in intermediate-band materials is intricately linked to solar concentration and étendue mismatch between absorption and emission. Generally, to achieve voltage matching under suboptimal conditions, an additional degree of freedom in the absorption thresholds of the material through a carrier relaxation or ratchet step is required. We explain why the ideal ratchet step decreases with solar concentration and how it depends on the radiative efficiency and emission étendue of the individual transitions and provide illustrative examples. The ideal ratchet step at 1 sun is largely given by the étendue mismatch between incoming sunlight and emitted light, which results in a step on the order of 270 meV that increases logarithmically with the ratio of luminescence-extraction efficiencies of the transitions. For solar cell concepts that use Auger-type carrier-carrier interactions or molecular triplet states for energetic up-conversion or down-conversion, the ideal band-gap combinations and achievable efficiencies also depend on interaction rates. We show that Auger-assisted solar cells suffer more strongly from finite interaction rates than carrier-multiplication devices.

## I. INTRODUCTION

Solar cells that are bounded by the Shockley-Queisser (SQ) efficiency limit possess one carrier temperature and nonequilibrium carrier populations in the valence band (VB) and conduction band (CB) only. To surpass the SQ efficiency limit, additional nonequilibrium carrier populations must be established, either through gradients in temperature or through multiple quasi-Fermi-level separations. These separate electron populations can be arranged in series or in parallel and can be mediated via luminescence or carrier interactions. The various permutations of series, parallel, and luminescent coupling are shown in Fig. 1.

The most familiar and most successful concept is the multijunction solar cell in the two-terminal configuration, where two or more materials with different band gaps are integrated in a single device and connected in series. A requirement for high efficiency is that the currents delivered at maximum power by each junction are equal [1]. A mild variation of this is to permit luminescent coupling between the subcells, where excess photogeneration in high-gap subcells can be transferred radiatively to lower-gap subcells. This effect does not increase the maximum efficiency attainable but serves as an example where an additional process can help recover power that would otherwise be lost, in this case decreasing the variability of the output under changeable illumination conditions [2] and the need for precise band-gap engineering [3,4]. Prospective multijunction solar cells that make use of this (two-terminal multijunction solar cells with luminescent coupling) require a high material quality and can be categorized as belonging to both the "series" category and the "luminescence" category. For completeness, we note the extreme case of a purely luminescent solar collector, the luminescent solar concentrator, which in the case of a single fluorescent sheet is bounded by the Shockley-Queisser limit but has potential merit for reducing the need for photovoltaic material [5]. It relies on frequency conversion of light impinging on luminophores and redirection to the edges of a waveguide, where it is collected by a conventional solar cell.

In this paper we are concerned with all systems enclosed within the parallel ellipse. The key to efficient operation of any parallel connected network of cells is to ensure that the free energies of the carrier populations, represented by quasi-Fermi-levels, match [6]. Any one of the schemes classified as parallel contains at least two well-separated absorption thresholds, each associated with a different quasi-Fermi-level separation (QFLS). When these transitions are represented in an equivalent circuit, any separation of quasi-Fermi-levels can be described as an internal voltage, and hence the term "voltage matching" has been coined as a measure of the internal free energy of carriers. Imperfect voltage matching leads to a loss of free energy for one of the parallel transitions compared with the optimum, just as imperfect current matching leads to a loss of current.

*andreas.pusch@gmx.net

An example of a material that can support multiple QFLS is the intermediate-band (IB) solar cell (IBSC), where a second energy gap is used to slow interband carrier equilibration [7] and thus allows excitation of carriers from an intermediate state or band to the CB by a sequential absorption process instead of thermal excitation via carrier-carrier or phonon-scattering processes. In this paper, we use the term "intermediate band" to describe both intermediate bands and intermediate states that do not form a band. The distinction is not important for the following discussion of voltage matching as long as quasi-thermal-equilibrium has been established such that the occupancy of the state or band can be described by a quasi-Fermi-level.

The limiting efficiencies for the different concepts shown in the Venn diagram in Fig. 1 were calculated in a number of previous publications. The concept of an IBSC, a device with increased photocurrent due to sequential absorption of below-band-gap photons in a single material, was first formulated as an impurity photovoltaic device, that uses impurity levels as stepping stones for sequential absorption via discrete electronic states by Wolf [8] and later revised by Luque and Martí [9] using intermediate bands. Yoshida *et al.* [10] showed that a ratchet or relaxation step ($\Delta E$) can increase the efficiency of an IBSC at low concentration, despite introducing a thermalization loss into the sequential-absorption mechanism. Two configurations for the band alignment of such a electronic ratchet are shown in Fig. 2, where the ratchet ($\Delta E$) is located in either the IB or the CB and the carrier populations defined in the conduction, valence, and intermediate bands are described by quasi-Fermi-levels $\mu_C$, $\mu_V$, and $\mu_I$, respectively. A positive ratchet step, corresponding to an exothermic process, ensures that the energy of the absorption thresholds—and more importantly the recombination thresholds—of the below-band-gap transitions add up to more than the absorption threshold of the VB-to-CB transition (i.e., $E_1 + E_2 > E_g$).

The ratchet can be implemented using several different phenomena [11], typically using a forbidden transition. This can proceed via physical separation of optical transitions [12–14] to form a "spatial ratchet," using spin-forbidden transitions as might be achievable in dilute magnetic semiconductors [15] or molecular triplet states [16] as a "spin ratchet," or by a separation of the carriers in momentum-space [17].

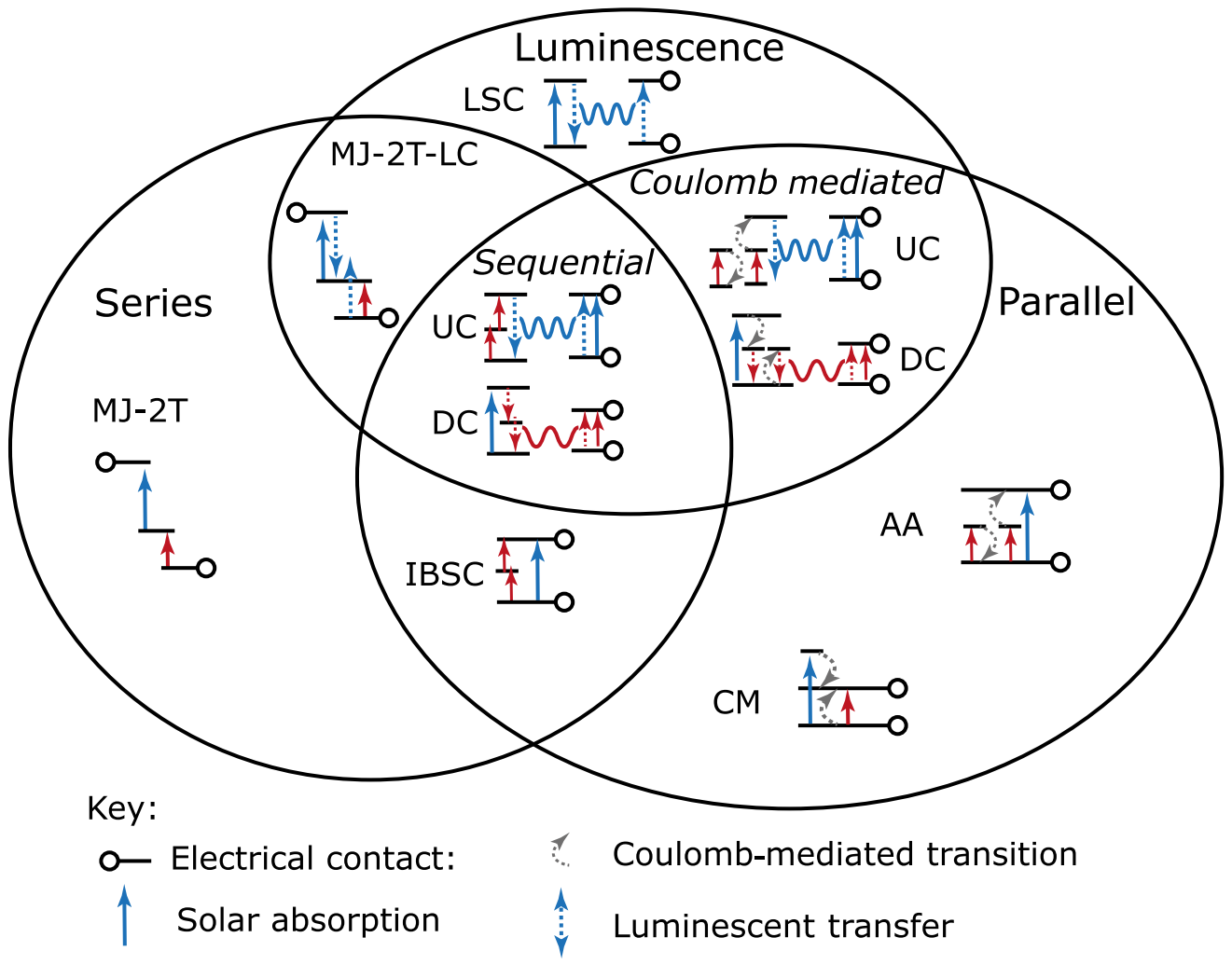

FIG. 1. A Venn diagram of the different advanced-concept solar cells; two-terminal multijunction with selective mirror (MJ-2T), two-terminal multijunction with luminescent coupling (MJ-2T-LC) through luminescent transfer of excitations, luminescent solar concentrator (LSC), up-conversion (UC) and down-conversion (DC) through sequential absorption or emission, or through Coulomb-mediated processes, intermediate-band solar cell (IBSC), Auger-assisted solar cell (AA), carrier-multiplication solar cell (CM) . Different elements can be connected in series, in parallel, or via luminescence. The concepts we focus on here are the ones that contain components that are connected in parallel.

In contrast, sequential-absorption up-conversion (UC) [18] and sequential-emission down-conversion (DC) [19] enable luminescent transfer of excitations by use of an IB material in front of or behind a single-junction solar cell. The sequential absorption process is sometimes termed "excited-state absorption" [20]. The importance of a relaxation step for UC to work efficiently at low-to-medium concentrations was recognized by Trupke *et al.* [18] and emphasized later in Ref. [21].

Coulomb-mediated carrier-carrier interactions also enable a separate class of up-conversion and down-conversion devices. In the case of up-conversion this process relies on two separate optical absorption events followed by an interaction between those excited states. Triplet-triplet annhiliation is such a process in molecular materials [16] and energy-transfer up-conversion is such a process in ionic systems [20]. Down-conversion is simply the reverse process, corresponding to singlet fission in molecular materials [22] and cooperative energy transfer in ionic systems [23].

The final class of parallel devices is represented by two Coulomb-mediated devices where the relevant bands are contacted directly, dispensing with the luminescent step altogether. Internal carrier multiplication (CM) [24,25], multiple-exciton generation [26], and direct singlet fission in molecular materials [27] correspond to systems where contacts are made to the lowest two energy bands in the system. The alternative is to contact the highest two energy bands, similarly to what is done in the IBSC, but distinct in that carriers are excited via an Auger-assisted (AA) process [28,29] or the molecular equivalent triplet-triplet annihilation [30], the latter having been demonstrated experimentally [31,32]. In all these instances, an overmatching of the QFLS between the IB and the VB and between the CB and the VB is required to provide a free-energy differential to drive the CM process, a requirement that has been articulated in the molecular-singlet-fission-solar cell concept [22].

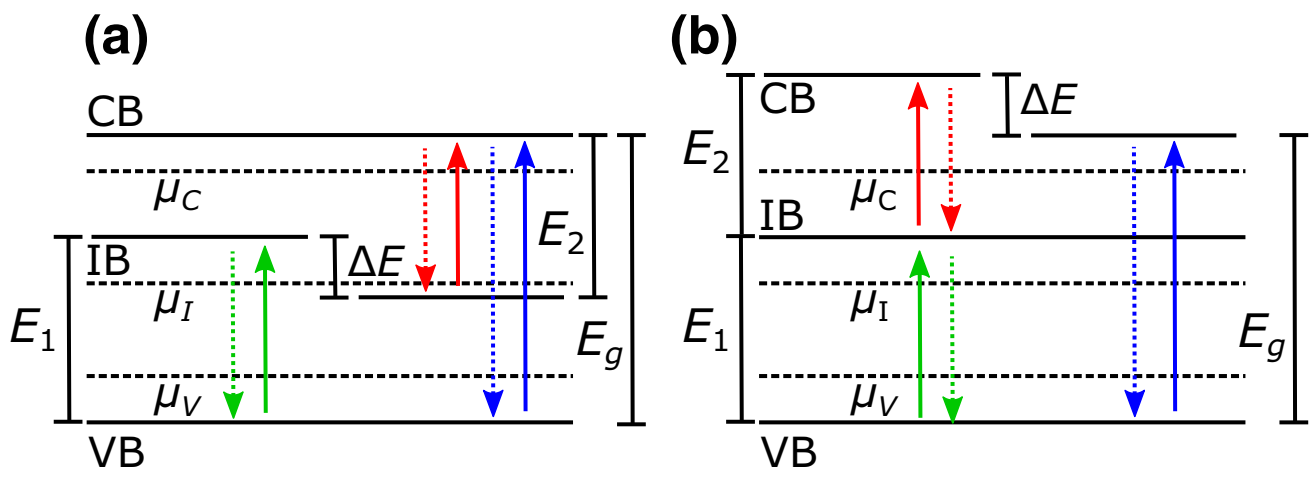

FIG. 2. An IB material with sequential absorption as a mechanism to promote carriers from the VB to the CB via the IB: (a) ratchet step in the IB: (b) ratchet step in the CB.

In this work, we show how the étendue of absorption and emission, radiative efficiencies, and interaction rates influence the size of the ratchet step that is necessary for parallel connected solar cells to attain their highest efficiency. We start with a thermodynamic description of an IB material in general and then consider the luminescent approaches, the purely parallel approach, especially under the consideration of finite Auger interaction rates, and finally the IBSC approach.

## II. OPEN-CIRCUIT QUASI-FERMI-LEVEL SEPARATIONS FOR BANDWIDTH-LIMITED TRANSITIONS

The concept of current matching in series-connected tandem solar cells stems from the intuitive concept of conservation of charge and current continuity. However in parallel-connected cells the concept of voltage matching is subtler. Driving each of the transitions with an incoming light field induces a QFLS between the two states involved in the transition. This QFLS is equivalent to a Gibbs free energy between the carrier populations in the respective bands; in a single-junction cell it corresponds to the potential for electrical work the system can perform and corresponds to the internal voltage of the device. With multiple transitions in a device, such as in the IBSC, the sum of the free energies in the parallel subunits needs to be matched for optimal operation, which, in the context of an equivalent circuit, corresponds to matching the internal voltages of the device.

Intermediate-band materials operate at open circuit in photonic UC and DC. Current is extracted from different bands for the CM and AA solar cells and the sequential-absorption-based IBSC. Current extraction means that optimal voltage matching should ideally occur at the maximum power point of the $I$-$V$ characteristic, which is slightly displaced from the open-circuit voltage. Nonetheless, the consideration of voltage matching at open circuit provides a good approximation.

To achieve voltage matching in the parallel class of solar cell concepts, it is necessary to have a degree of freedom to choose the absorption thresholds. As we explain in detail below, the Gibbs free energy of an electron-hole pair in a transition driven by sunlight is a result of étendue mismatch and radiative efficiency. At low concentration, all the parallel concepts reach maximum efficiency only if a ratchet or relaxation step is introduced such that the sum of the recombination threshold energies of the below-bandgap transitions does not correspond to the energy of the recombination threshold for the VB-to-CB transition (i.e., $E_1 + E_2 \neq E_g$).

We now analyze the open-circuit voltage of an intermediate-band material—which could be used to implement either of the parallel schemes—with the aim of elucidating the relationship between étendue, radiative efficiency, and the magnitude of the ratchet step. We make the Boltzmann approximation for both the illumination from the Sun and the luminescence from the material to illustrate the fundamental thermodynamic considerations involved in parallel solar cells. In the Boltzmann approximation the maximum Gibbs free energy of an electron-hole pair in an absorber with sharp threshold $E_g$ at temperature $T_c$ under illumination by a blackbody of temperature $T_s$ can be written as [33,34]

$$F = E_g\left(1 - \frac{T_c}{T_s}\right) + kT_c \ln\left(\frac{\Omega_{\rm abs}}{\Omega_{\rm em}}\frac{\gamma(E_g, T_s)}{\gamma(E_g, T_c)}\eta_{\rm ext}\right), \quad (1)$$

where $\Omega_{\rm abs}$ is the étendue of the absorbed sunlight and $\Omega_{\rm em}$ is the étendue of the luminescence emitted by the absorber. At full concentration the étendues are equal, while there is a factor of approximately 46 260 between them at 1 sun. The multiplier $\gamma(E_g, T)$ is given by

$$\gamma(E_g, T) = T\left(E_g^2 + 2TE_g + 2T^2\right). \quad (2)$$

To take into account a finite absorption bandwidth we can replace $\gamma(E_g, T)$ of a step absorber with a function $\gamma_l(E_1, E_2, T)$ that incorporates lower and upper absorption limits $E_1$ and $E_2$ in each band as

$$\gamma_l(E_1, E_2, T) = \gamma(E_1, T) - e^{(E_1-E_2)/kT}\gamma(E_2, T). \quad (3)$$

The Boltzmann approximation for the incoming sunlight becomes inaccurate at small band gaps and should not be used in quantitative modeling of small-band-gap absorbers. Nonetheless, as the free energy depends only logarithmically on the generation rate, these simple equations for the achievable free energy under sunlight illumination remain useful to illustrate the operation principles of parallel solar cells.

Consider a band arrangement with sequential absorption and emission processes as in Fig. 2. There are three absorption thresholds, which can result in three sets of states with their own quasi-Fermi-level under illumination. For the carrier free energies in the $\mu_{ij} = \mu_i - \mu_j$ or QFLSs, between valence ($V$), intermediate ($I$) and conduction ($C$) states, with

$$\mu_{IV} + \mu_{CI} = \mu_{CV}. \quad (4)$$

Using the QFLSs of the individual transitions, we obtain

$$\mu_{IV} = E_1\left(1 - \frac{T_c}{T_s}\right) + kT_c \ln\left(\frac{\Omega_{\rm abs}}{\Omega_{\rm em}}\frac{\gamma_l(E_1, E_2, T_s)}{\gamma_l(E_1, E_2, T_c)}\eta_{\rm ext}^{IV}\right), \quad (5)$$

$$\mu_{CI} = E_2\left(1 - \frac{T_c}{T_s}\right) + kT_c \ln\left(\frac{\Omega_{\rm abs}}{\Omega_{\rm em}}\frac{\gamma_l(E_2, E_g, T_s)}{\gamma_l(E_2, E_g, T_c)}\eta_{\rm ext}^{CI}\right), \quad (6)$$

and

$$\mu_{CV} = E_g\left(1 - \frac{T_c}{T_s}\right) + kT_c \ln\left(\frac{\Omega_{\rm abs}}{\Omega_{\rm em}}\frac{\gamma(E_g, T_s)}{\gamma(E_g, T_c)}\eta_{\rm ext}^{CV}\right), \quad (7)$$

where $\mu_{CV}$ is the QFLS of an absorber with external luminescence extraction efficiency $\eta_{\rm ext}^{CV}$ in the absence of intermediate states.

The ratchet step $\Delta E^{\rm match}$ that leads to a matching at open circuit of the free energies between the two low-band-gap absorbers in series and the high-band-gap absorber is given by the implicit condition

$$W_{\rm OC}(E_g) = W_{\rm OC}(E_g + \Delta E^{\rm match} - E_1) + W_{\rm OC}(E_1), \quad (8)$$

where $W_{\rm OC}(E) = E - qV_{\rm OC}$ is the difference between the absorption threshold and the open-circuit voltage of a solar cell. To understand the dependencies of the ratchet step we can approximate this with the explicit relation for the ratchet step that matches the voltages at open circuit:

$$\Delta E^{\rm match} \approx \frac{kT_cT_s}{T_s - T_c}\ln\left(\frac{\Omega_{\rm em}}{\Omega_{\rm abs}}\frac{\eta_{\rm ext}^{CV}}{\eta_{\rm ext}^{CI}\eta_{\rm ext}^{IV}}\frac{\gamma_l(E_1, E_2, T_c)\gamma_l(E_2, E_g, T_c)\gamma(E_g, T_s)}{\gamma_l(E_1, E_2, T_s)\gamma_l(E_2, E_g, T_s)\gamma(E_g, T_c)}\right). \quad (9)$$

To analyze the voltage-matched ratchet step it is instructive to look first at the ratios of $\gamma_l(E_i, E_j, T_s)$ to $\gamma_l(E_i, E_j, T_c)$. For the absorption thresholds relevant for solar cells, this ratio is on the order of $T_s/T_c \approx 20$ and thus results in a negative contribution to the ratchet step of approximately $3kT_c$, or approximately 75 meV. The contribution of the ratio between emission and absorption étendues $\Omega_{\rm em}/\Omega_{\rm abs}$ depends on the concentration of the incoming sunlight but is usually much greater than the contribution from the temperature ratio. For unconcentrated sunlight and a solar cell with a back reflector, it is approximately $11kT_c$, or approximately 278 meV. The ideal ratchet step for the ideal IBSC derived numerically from the current-voltage characteristics of the device [10] can thus be estimated from basic thermodynamic considerations. Note that this is an estimate based on matching the open-circuit voltage, while the relevant voltage in the IBSC is the maximum-power-point voltage, a point we return to in Sec. III C. Finally, the contribution of the external luminescence extraction efficiencies $\eta_{\rm ext}$ depends on the material quality and optical geometry (i.e., light trapping). It can easily reach a magnitude similar to the étendue mismatch.

## III. APPLICATION TO PARALLEL CLASS OF SOLAR CELLS

In the illustrative examples of the impact of ratchet steps and voltage matching for different types of parallel solar cells presented in this section, we approximate the incoming sunlight as blackbody radiation with a temperature of 6000 K to simplify the calculations while retaining the important physics of the problem.

### A. Photonic up-conversion and down-conversion

In a photonic UC solar cell, a material with intermediate states that act as a stepping stone for the up-conversion process is added at the front or back of a conventional single-junction solar cell. Adding it at the back is usually more efficient. The down-converter must be placed in front of a conventional solar cell and ideally converts one high-energy photon into two low-energy photons with energy just above the band gap of the conventional solar cell. It is therefore advantageous if both low-energy transitions occur at the same energy.

The photonic up-converting or down-converting materials are not electrically contacted, and hence they operate at open circuit and transfer power radiatively, as illustrated in Fig. 3. From the condition (9) on the ratchet step $\Delta E$, we can see whether a particular system is going to act as an up-converter or a down-converter. If the sum of the QFLSs that can be sustained by the two low-energy transitions is smaller than the QFLS sustained by the high-energy transition, down-conversion will predominantly occur. The recombination through the high-energy transition is suppressed compared with recombination through the low-energy transitions. If, however, the QFLS of the high-energy transition is smaller, up-conversion will occur. An up-converter with a below-optimal ratchet step $\Delta E$ shows a quadratic dependence of the UC intensity on the intensity of the low-energy light, while it will operate in the linear regime if $\Delta E$ is larger than optimal.

We illustrate this for the case of a symmetric sequential-absorption up-converter and a down-converter. In the symmetric case, there are only two absorption thresholds, at $E_1$ and $E_g$, and light with energies between $E_1$ and $E_g$ is absorbed equally by the VB-to-IB transition and the IB-to-CB transition. If the up-converter is placed behind an optically thick solar cell of band gap $E_g^c < E_g$, no sunlight will excite the VB-to-CB transition in the up-converter.

At open circuit, no current will flow in either of the bands, so we can write

$$G_{VC} + G_{IC} = R_{CV} + R_{CI} \tag{10}$$

to balance the net generation in the conduction band.

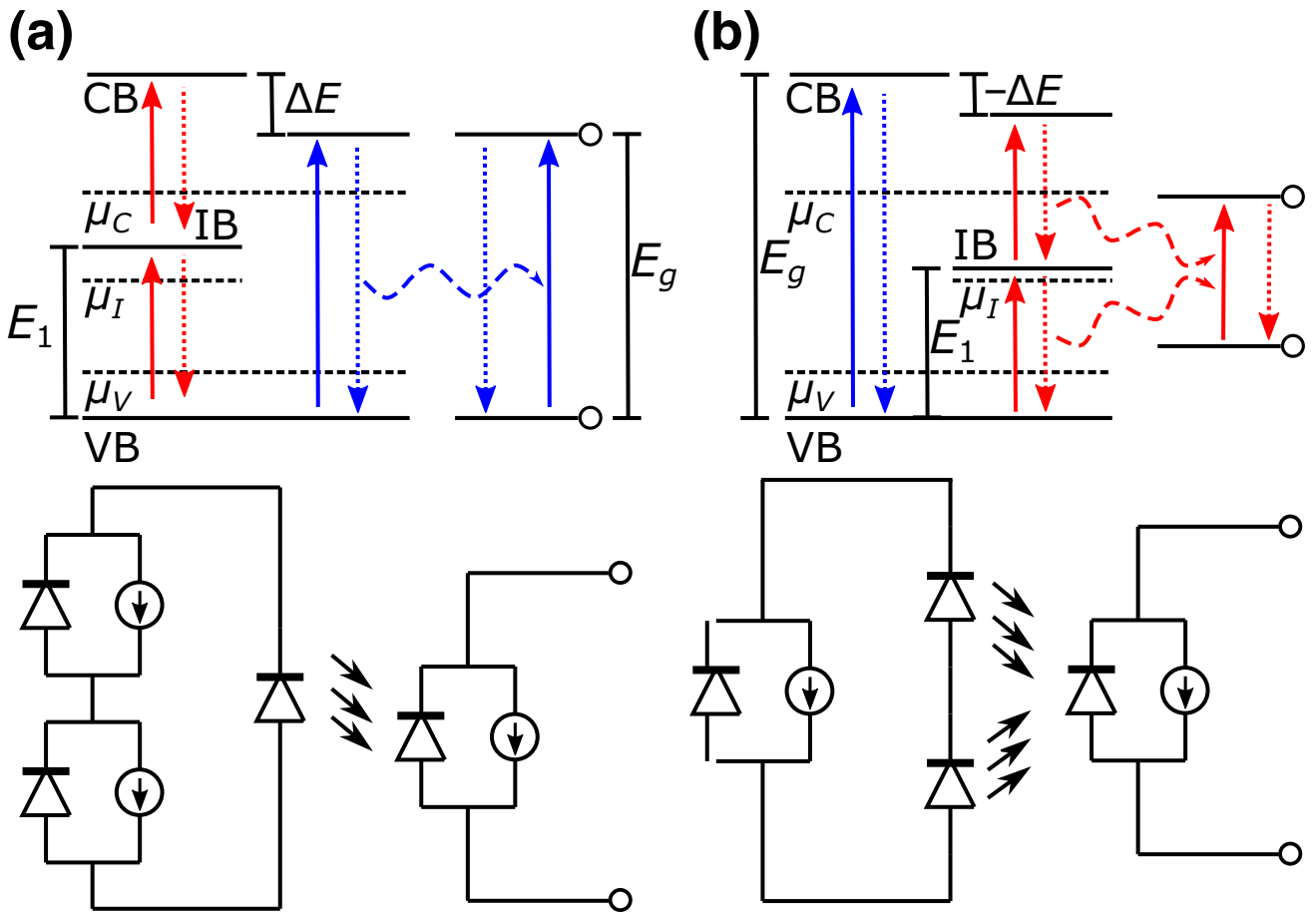

FIG. 3. Band alignment and processes in (a) a symmetric sequential-absorption up-converter and (b) a sequential-emission down-converter and their equivalent circuits.

The balance in the intermediate states,

$$G_{VI} - G_{IC} = R_{IV} - R_{CI}, \tag{11}$$

needs to be considered only in the asymmetric case as it is automatically fulfilled in the symmetric case. $G_{ij}$ and $R_{ij}$ denote the generation and recombination rates from band $i$ to band $j$, respectively. Since $\mu_{IV} + \mu_{CI} = \mu_{CV}$, we obtain the quasi-Fermi-level separation between the valence band and the conduction band for the symmetric case with equal absorption and recombination rates and assuming ideality factor 1 for each transition:

$$\mu_{CV} = 2\mu_{IV} = 2\ln\left(\frac{\sqrt{(R_{IV}^0)^2 + 4R_{CV}^0(G_{VC} + G_{VI})} - R_{IV}^0}{2R_{CV}^0}\right). \tag{12}$$

The generation rates are given by

$$G(E_{\min}, E_{\max}) = \frac{2\pi f}{(2\pi\hbar)^3 c^2}\int_{E_{\min}}^{E_{\max}} \frac{a(E)E^2 dE}{\exp\left(\frac{E}{kT_S}\right) - 1}, \tag{13}$$

with $f = 1$ corresponding to full concentration, $f = 1/46266$ corresponding to 1 sun, and $T_S$ being the temperature of the Sun. $E_{\min}$ and $E_{\max}$ are lower and upper absorption boundaries, respectively, and the absorptivity $a$ is maximally 0.5 for the symmetric intermediate states as the transitions into and out of the state share the optical excitation.

Radiative recombination can usually be calculated in the Boltzmann approximation and depends on the optical density of states as well as electronic properties. For emission across the surface to air from a material with a sharp absorption edge $E_{\min}$ and an optional upper absorption edge $E_{\max}$, absorptivity $a$, and QFLS $\mu$ we get

$$R^{\text{air}}(E_{\min}, E_{\max}, a) = ae^{\mu/kT_c}\left(e^{-E_{\min}/kT_c}\gamma(E_{\min}, T_c) - e^{-E_{\max}/kT_c}\gamma(E_{\max}, T_c)\right). \tag{14}$$

Here the étendue of emission $\Omega_{\text{em}}$ is given by $2\pi$. Luminescent transfer usually occurs between materials with a refractive index larger than unity. For optically thick materials the rate of emission and therefore also $\Omega_{\text{em}}$ is multiplied by a factor of $n^2T$, where $n^2$ is the minimum of the refractive indices and $T$ is the transmission across the interface. If the material is not optically thick, the relative emission étendue can go up to $4n^2$ due to the properties of radiative transfer at large angles [4]. If optical coupling is mediated by resonant processes or near-field coupling, the density of optical states, and with it the étendue of a particular transition, can be enhanced by orders of magnitude [35]. This also means that the voltage-matching condition that enables UC or DC of the incoming photon flux depends not only on the concentration of the incoming sunlight but also on the optical density of states at the emission frequencies.

As an illustrative example of the considerations above, we show results for the concentration-dependent additional photocurrent that can be achieved in a CdTe solar cell if a symmetric sequential-absorption up-converter is placed between the single-junction CdTe solar cell and a back mirror [see Fig. 4(a)]. In the calculations we ignore nonradiative recombination in the up-converter, we assume optically thick materials, and we also ignore radiative recombination from the CdTe cell that illuminates the up-converting material as recombination in CdTe is dominated by nonradiative processes. Therefore,

$$R_{IV}^{\mathrm{rad,UC}} = R_{CI}^{\mathrm{UC}} = R^{\mathrm{air}}(E_1, E_g, 0.5) \tag{15}$$

and

$$R_{CV}^{\mathrm{rad,UC}} = R^{\mathrm{air}}(E_g, \infty, 1)n^2. \tag{16}$$

The étendue of up-converted emission is enhanced by the square of the refractive index of the up-converting material (as long as it is below the refractive index $n_{\mathrm{CdTe}} = 3$), while the étendue of the emission below the solar cell band gap is given by the emission étendue of the front surface of the cell as this emission exits through the front surface between the solar cell and air. With a higher refractive index, the UC therefore reaches its peak efficiency at lower concentration than with a lower refractive index.

A larger ratchet step leads to a high UC efficiency at low solar concentrations. At high solar concentrations, the UC flux is proportional to the incoming photon flux (i.e. the UC becomes a process linear in intensity). The smaller ratchet step allows only low UC efficiency at low concentration, yet the UC efficiency rises with concentration. Ultimately, a higher photon flux can be reached at high concentration, compared with the larger ratchet step, as more of the below-band-gap spectrum is absorbed by the up-converter.

The down-converter is placed in front of the single-junction solar cell. For our illustrative results we assume a silicon solar cell and a down-converter that has a narrow low-energy emission band with a width of $kT_c$, which minimizes losses due to absorption of light that cannot be down-converted. This means that the luminescence from the transitions is given by

$$R_{IV}^{\mathrm{rad,DC}} = R_{CI}^{\mathrm{DC}} = R^{\mathrm{air}}(E_1, E_1 + T_c, 0.5)(n^2 + 1) \tag{17}$$

and

$$R_{CV}^{\mathrm{rad,DC}} = R^{\mathrm{air}}(E_g, \infty, 1)(n^2 + 1). \tag{18}$$

The beneficial recombination is only the part emitted toward the cell (represented by $n^2$); the part emitted toward air (represented by 1) is lost. In a functioning down-converter, $R_{CV}$ will be negligible compared with $R_{IV}$.

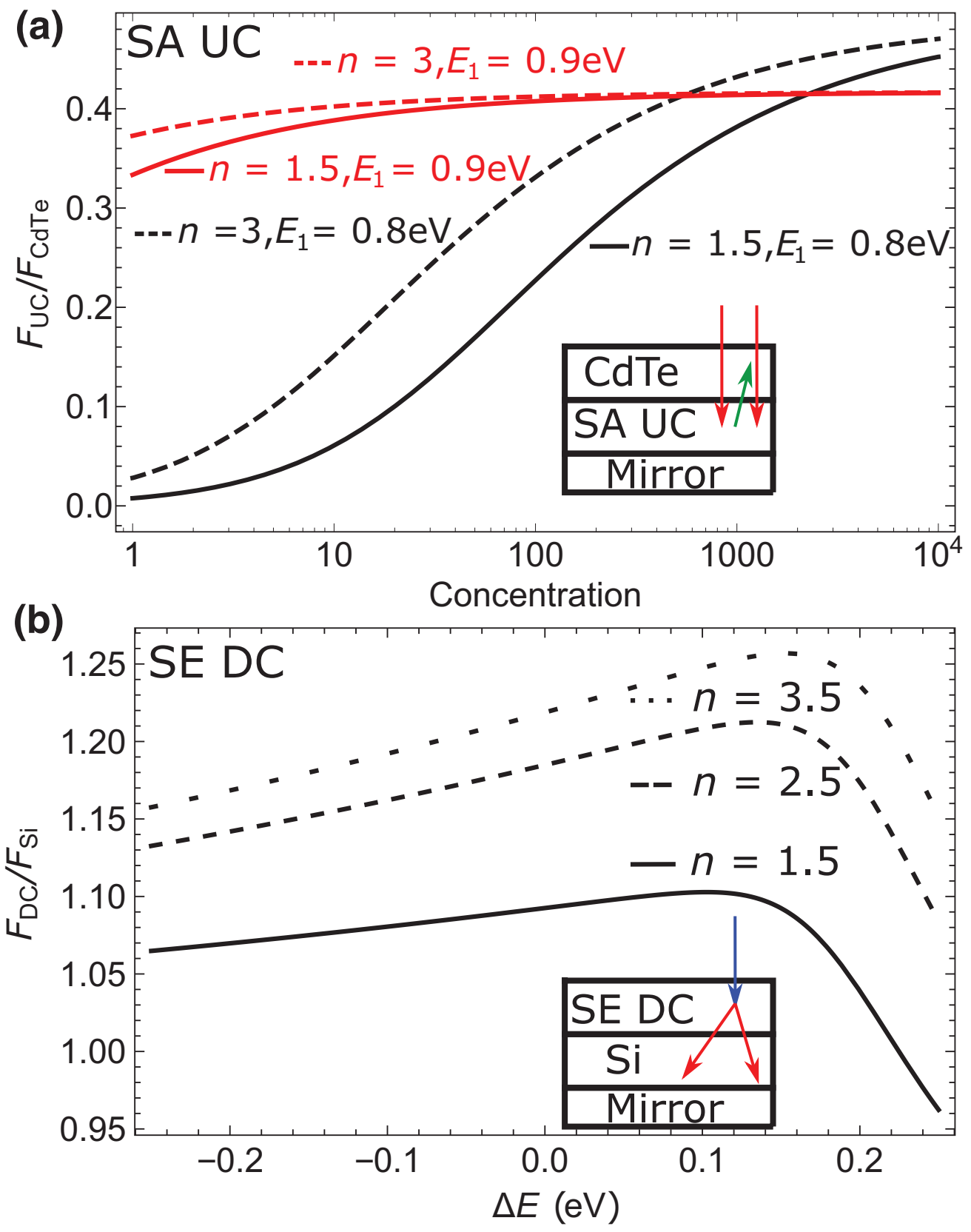

FIG. 4. (a) The additional photon flux $F_{\mathrm{UC}}$ onto a CdTe solar cell with $E_g = 1.55$ for a symmetric up-converter placed behind the cell with UC emission at 1.55 eV and an absorption threshold of $E_1$. $F_{\mathrm{UC}}$ is normalized to the incoming flux above 1.55 eV ($I_{\mathrm{CdTe}}$). It therefore represents the potential for a relative increase in photocurrent. The inset shows the geometrical configuration. The étendue for the emission into the solar cell is given by the refractive index of the up-converting material, while the étendue for the emission below the band gap of the solar cell is given by the étendue for emission from the solar cell, where total internal reflection occurs for large angles. (b) The total photon flux $F_{\mathrm{DC}}$ from a down-converting material onto a Si solar cell, normalized to the incoming photon flux from the Sun above 1.12 eV ($F_{\mathrm{Si}}$) and plotted against the ratchet step $\Delta E$ in the down-converter ($E_g = 2E_1 - \Delta E$). The DC emission threshold $E_1$ is set to 1.2 eV to match the onset of strong absorption in a typical Si solar cell and the emission bandwidth is set to a small energy interval of $kT_c$ to minimize the absorption by the down-converter. The inset shows the placement of the down-converter in front of the cell. SA sequential absorption; SE, sequential emission.

Thus, a high refractive index is beneficial for two reasons. Firstly, it increases the fraction of emission that is directed toward the solar cell compared with the emission directed toward air. In Fig. 4(b), showing exemplary results for a silicon solar cell, we see that a sequential-emission down-converter with refractive index $n = 1.5$ increases the photon flux hitting the solar cell only minimally compared with the incoming photon flux from the Sun as too much is emitted from the front of the cell. When the refractive index is increased, the photon flux on the solar cell can be increased compared with the incoming photon flux and the down-converter is useful. Varying the ratchet step from negative values (the intermediate states have a finite bandwidth and carriers relax to the bottom before emitting light again) through to positive values, we see how the DC efficiency increases slightly as more of the spectrum is multiplied.

Positive values of the ratchet step for DC have previously often been excluded from the analysis (e.g., in the original work by Trupke *et al.* [19]). A positive value of the ratchet step corresponds to an endothermic process and is not forbidden as long as thermal equilibration within a band occurs. When the ratchet step becomes too large, the down-converter stops down-converting (i.e., the open-circuit voltage in the down-converter now starts to favor luminescence from the upper transition) and the DC efficiency drops drastically.

These results also reveal the second reason why a higher refractive index can be beneficial for DC: the matching of the open-circuit voltages is shifted toward higher concentration for a higher refractive index. This allows a slightly larger ratchet step in the down-converter and therefore a higher current. It occurs, because in the down-converter, both the low-energy transitions and the high-energy transitions emit into a higher étendue if the refractive index is increased, unlike in the up-converter, where only the above-band-gap transitions emit into a higher étendue.

An additional consideration for photonic UC or DC is the external luminescence extraction efficiency of the different transitions. For UC to be efficient, the CB-to-VB transition $\eta_{\mathrm{ext}}^{CV}$ has to be close to unity, otherwise most of the carriers will be lost to nonradiative recombination with

$$R_{CV}^{\mathrm{nonrad}} = R_{CV}^{\mathrm{rad}}(1/\eta_{\mathrm{ext}}^{CV} - 1). \tag{19}$$

In contrast, some nonradiative recombination in the IB can be balanced by an increased ratchet step without compromising the efficiency of the up-conversion process, although this will decrease the number of low-energy photons that can be harnessed. In the DC case, the low-energy transitions must be radiatively efficient, otherwise it is not possible to achieve the desired above-unity external quantum efficiency for the high-energy photons. A nonradiative high-energy transition can, however, be tolerated in a down-converter.

The considerations presented here also apply for molecular UC and DC via triplet-triplet annihilation or singlet fission in organic materials or UC and DC using lanthanides. Specific consequences of exchanging sequential absorption for a Coulomb-mediated interaction process are discussed in the following section with emphasis on direct injection instead of luminescent transport.

### B. Purely parallel solar cells

The CM solar cell and the AA solar cell are distinguished in their principle of operation mainly by the placement of the contacts. In an AA cell, the contacts are placed at the valence band and the conduction band, while the CM cell has to be contacted in the intermediate band and care has to be taken that the carriers in the CB cannot simply reach the contacts without first relaxing to the IB via Coulomb-mediated carrier multiplication. The AA cell is therefore a high-voltage, low-current cell and the CM cell is a low-voltage, high-current cell. Figure 5 illustrates how a Coulomb-mediated process enables carrier up-conversion and carrier multiplication and presents an equivalent-circuit diagram that represents both devices. In the AA cell, the ratchet step represents carrier thermalization on Auger recombination, and a positive ratchet step results in an exothermic reaction of two carriers in the IB scattering into a carrier in the VB and the CB, respectively (i.e., heat is supplied to the lattice on Auger recombination). A positive ratchet step in the CM cell thus represents an endothermic reaction and the lattice has to provide heat to enable carrier multiplication [22].

One can switch from one cell type to the other by exchanging which transition is electrically contacted to and thus represented by the diode. The exact nature of the voltage-dependent current source is illustrated using the results that follow.

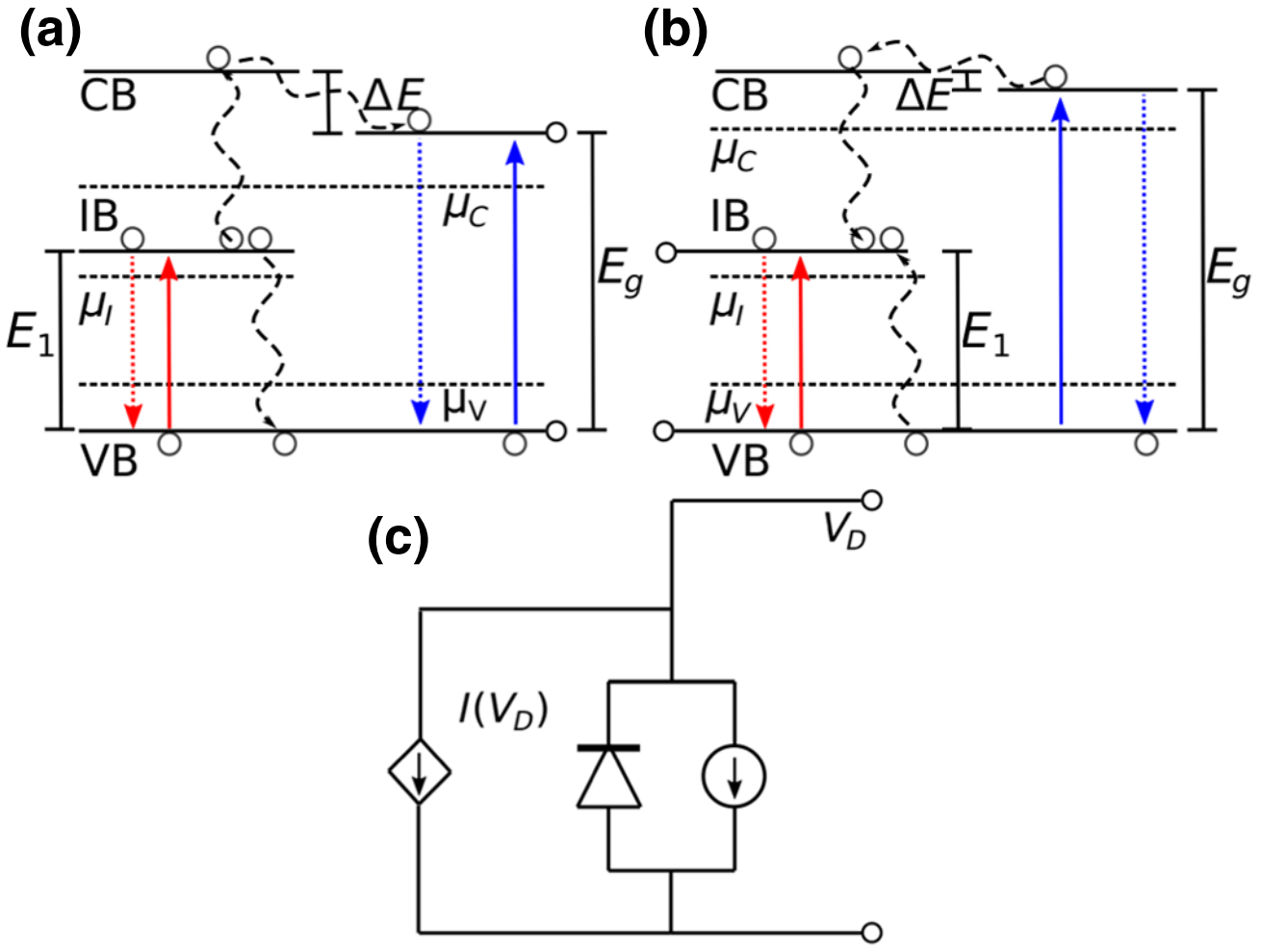

FIG. 5. Band alignment, absorption process, and Coulomb-mediated carrier-carrier interaction process in (a) an AA solar cell and (b) a CM solar cell. (c) Equivalent circuit of an AA cell and a CM cell. For the AA cell the diode represents the high-band-gap transition and the IB is represented by an additional voltage-dependent current source. For the CM cell, the diode represents the low-band-gap transition and the CB is represented by a voltage-dependent power source.

Both cells require strong Auger-type carrier-carrier interactions in the intermediate band to function. Note that interactions that allow carriers to also relax from the CB to the IB without Coulomb-mediated multiplication lead to significantly-less-efficient carrier multiplication and can render the AA solar cell completely ineffective due to strongly enhanced nonradiative recombination.

The Auger process and its inverse, the impact-ionization process, are depicted in Fig. 5(c). The rate of the Auger process $r_A$ depends on the carrier-carrier-interaction matrix elements $M_{ijkl}$ between the initial intermediate states $i$ and $j$ and the final states $k$ and $l$ in the valence band and the conduction band respectively, as well as their occupations $f_n$:

$$r_A = \sum_{ijkl} M_{ijkl} f_i f_j (1-f_k)(1-f_l). \tag{20}$$

Impact ionization, as the reverse process, depends inversely on the occupation probabilities, resulting in

$$r_{II} = \sum_{ijkl} M_{ijkl} (1-f_i)(1-f_j) f_k f_l. \tag{21}$$

While the interaction matrix elements are given by Coulomb integrals over the electronic wave functions and are largely determined by the material itself, the occupations depend on external conditions such as illumination and applied voltage. Therefore, we aggregate the sum over the matrix elements into an interaction strength.

Under solar illumination it is usually safe to assume there is a quasiequilibrium between states without energy gaps, as intraband carrier-carrier interaction and carrier-phonon interaction are much faster than interband and radiative processes. This allows us to rewrite the occupation probabilites as

$$\begin{aligned} f_n^{\rm VB} &= \frac{1}{e^{\frac{E_n-\mu_V}{kT_c}}+1}, \\ f_n^{\rm IB} &= \frac{1}{e^{\frac{E_n-\mu_I}{kT_c}}+1}, \\ f_n^{\rm CB} &= \frac{1}{e^{\frac{E_n-\mu_C}{kT_c}}+1}. \end{aligned} \tag{22}$$

For simplicity, we assume that the bandwidth of the intermediate states is small compared with the gaps between the bands. As a consequence, the energy difference between the two states in the VB and the CB that are connected by a Coulomb-mediated interaction event is given approximately by $2E_1 = E_g + dE$. Without restricting the validity of our results, we set the energy of the carrier in the valence band to 0. Finally, we use the Boltzmann approximation for VB and CB occupations, which is valid under sunlight illumination away from full concentration, to obtain

$$\begin{aligned} (1-f^{\rm VB}) &\approx e^{-\mu_V/kT_c}, \\ f^{\rm CB} &\approx e^{\mu_C - 2E_1/kT_c}. \end{aligned} \tag{23}$$

We then write the Auger up-conversion and impact-ionization rates as

$$\begin{aligned} r_A &= M\left(e^{\mu_I/kT_c} + e^{E_1/kT_c}\right)^{-2} e^{(2\mu_I-\mu_V)/kT_c}, \\ r_{II} &= M\left(e^{\mu_I/kT_c} + e^{E_1/kT_c}\right)^{-2} e^{\mu_C/kT_c}. \end{aligned} \tag{24}$$

The rates are equal if the sum of the chemical potentials (or Gibbs free energies) of the electrons in the initial states $2\mu_I$ equals the sum of the chemical potentials of the electrons in the final states $\mu_V + \mu_C$ as demanded by detailed balance.

To obtain a current-voltage relation we consider the metallic-intermediate-band case [36] in which the equilibrium Fermi level of the material is in the intermediate band. This means that $\mu_I$ is approximately fixed at $E_1$ under illumination, so we can replace $\mu_I$ with $E_1$ in Eq. (24).

The current-voltage characteristic is then given by

$$\begin{aligned} I^{\mathrm{AA},m}/q &= G_{VC} + r_{\rm II} - R_{CV} - r_A \\ &= G_{VC} - e^{\mu_{CV}/kT_c} R^0_{CV} + \frac{M}{4} e^{-E_1/kT_c} \\ &\quad \times \left(e^{\mu_{IV}/kT_c} - e^{(\mu_{CV}-\mu_{IV})/kT_c}\right) \end{aligned} \tag{25}$$

for the AA cell and by

$$\begin{aligned} I^{\mathrm{CM},m}/q &= G_{VI} - 2r_{\rm II} - R_{CV} + 2r_A \\ &= G_{VI} - e^{\mu_{CV}/kT_c} R^0_{CV} - \frac{M}{2} e^{-E_1/kT_c} \\ &\quad \times \left(e^{\mu_{IV}/kT_c} - e^{(\mu_{CV}-\mu_{IV})/kT_c}\right) \end{aligned} \tag{26}$$

for the CM cell.

The power extracted from the device can be written as

$$P^{\rm AA} = I^{\rm AA} V^{\rm AA} = I^{\rm AA} \mu_{CV}/q \tag{27}$$

for the AA cell and

$$P^{\rm CM} = I^{\rm CM} V^{\rm CM} = I^{\rm CM} \mu_{IV}/q \tag{28}$$

for the CM cell.

We need to solve for the QFLS of the transition not electrically contacted to (i.e., not fixed by) the applied voltage as a function of the voltage applied at the other transition, the Auger interaction rate $r^0_A$, carrier-generation rates $G_{VI}$ and $G_{VC}$, and recombination rates $R^0_{IV}$ and $R^0_{CV}$. The maximum power point (in dependence of the absorption thresholds $E_1$ and $E_g$ and the Auger rate $r_A^0$) can be found through a numerical root-finding algorithm. $r_A^0$ depends on the intermediate-state offset as $e^{-2E_1/T_c}$ for the intrinsic case and as $e^{-E_1/T_c}$ for the metallic IB case.

The limiting efficiencies for the AA cell and the CM cell are obtained by our assuming infinitely fast Auger interaction rates and instant carrier thermalization in each band, in addition to the usual SQ assumptions of sharp band edges $E_1$ for VB-to-IB transitions and $E_g$ for VB-to-CB transitions, infinite carrier mobility, absence of nonradiative recombination, and unity absorptivities. Here we look at the maximum efficiencies as a function of a finite Auger interaction strength, keeping the other idealizations. To provide intuition for the order of magnitude of the Auger interaction strengths and to provide a realistic band-gap dependence of the Auger interaction, we assume a semiconductor material that follows the Kane rule for the dependence of the electron effective mass on the band gap:

$$m_e(E_g) = \frac{m_0}{1 + 20eV/E_g}. \tag{29}$$

We take a typical bulk-semiconductor Auger interaction strength of $10^{-30}\mathrm{cm}^6/\mathrm{s}$. The maximum power point for specific band-gap combinations can be obtained numerically by taking the derivative of the power with respect to the applied voltage. We show the maximum efficiencies, the ideal band gaps, the relaxation step, and the QFLSs at the operating point as a function of Auger interaction strengths in Fig. 6 for both the CM solar cell and the AA solar cell with a metallic IB. In these plots the Auger rates are parametrized through the effective thickness of a bulk semiconductor that would deliver this Auger rate if it follows Eq. (29) for the IB effective density of states. The limiting efficiency is reached only for unrealistically thick devices, corresponding to unrealistically fast Auger interaction rates in bulk semiconductors.

Usually, it is assumed that the CM solar cell can double the current only upward from an energy above twice the fundamental band gap. We thus need to explain how it is possible to have a positive ratchet step for the CM cell that leads to $E_g < 2E_1$. Because it is indeed correct that individual carriers can undergo impact ionization only if they have twice the energy of the states in the IB, it seems natural to assume that $E_g \geq E_1$. Nonetheless, this is not a necessary condition. Consider an Auger interaction rate that is infinitely fast, which is the assumption made to obtain the limiting efficiency. In that case any finite thermal carrier occupation in above-band-gap states with $E = 2E_1$ is sufficient to obtain efficient carrier multiplication as the carriers in the above-band-gap states can multiply by drawing thermal energy from the lattice. This is the semiconductor equivalent to an endothermic chemical reaction and it is preferable for high efficiency if the interaction rate is not the limiting factor since it leads to a higher current overall. The endothermic energy step is reduced when finite Auger interaction rates are considered, and for slow Auger interaction, an exothermic process becomes necessary to efficiently drive carrier multiplication, as shown in Fig. 6(c). Experiments that find appreciable carrier multiplication only at excitation energies far above $2E_g$ [37,38] probably suffer from a combination of slow Auger interaction rates and competing relaxation pathways.

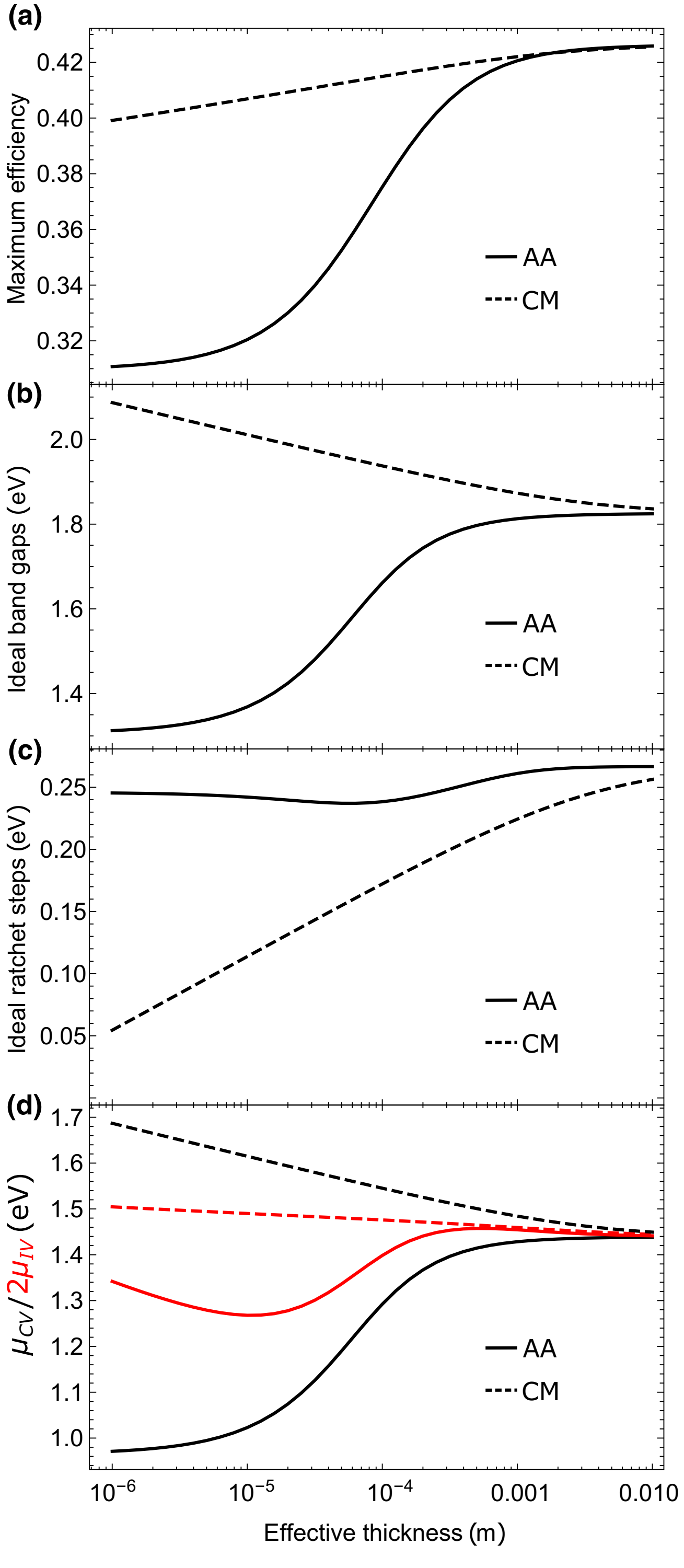

FIG. 6. (a) Maximally achievable efficiencies, (b) ideal band gaps, and (c) Ideal ratchet steps of AA (solid lines) and CM (dashed lines) solar cells as function of the effective Auger interaction thickness of the device. (d) The QFLSs of the individual transitions at the operating point. The lower of the transitions is multiplied by 2 to illustrate the loss in free energy on carrier multiplication for the CM solar cell and on Auger-mediated up-conversion for the AA solar cell.

We can clearly see from Fig. 6(a) that the CM cell is more robust to slow Auger interaction, making it the more realistic to achieve compared with the semiconductor implementation of an AA cell, provided selective contacts that allow only the carriers in the IB states to flow to the contacts, not carriers in the CB. The best ratchet steps shown in Fig. 6(c) show a logarithmic decrease of the ratchet step with the interaction rate for the CM device. This occurs because the driving force for the CM process, which is proportional to the quasi-Fermi-level of the conduction-band states [see Eq. (24)], gets stronger with decreasing ratchet step, which can compensate for the slower interaction rate. By symmetry, one could expect that the ideal ratchet step for the Auger device would increase to compensate for slower interaction rates, yet this is not the case. A close look at Eq. (24) reveals that a larger ratchet step does not increase the net Auger up-conversion rate, and it is therefore not helpful to increase the ratchet step beyond the voltage-matching requirement.

The QFLSs [see Fig. 6(d)] at the operating point conform to the requirement that the Gibbs free energy created in the process of Auger up-conversion or carrier multiplication has to be positive in the respective schemes. The stronger the Auger interaction, the smaller the free-energy differential that is necessary to drive the transitions efficiently.

Finally, in Figs. (7) and (8) we present the typical $I$-$V$ curves of the two devices, together with the quasi-Fermi-level separations of the bands that are not contacted, which allow us to understand whether Auger recombination or impact ionization dominates and also show that QFLSs at short circuit are a necessary feature in this type of device. Clearly, the CM cell has a substantial quasi-Fermi-level separation between the VB and the CB $\mu_{CV}$, even at short circuit, which enables a doubling of the current obtained from photons with energy above the VB-to-CB band gap $E_g$. This is necessary to efficiently drive the impact-ionization process. The QFLS $\mu_{CV}$ of the device increases toward the operating point, where it meets the point of twice the applied voltage. Here Auger recombination starts to dominate over carrier multiplication, and carrier multiplication ceases to function. Increasing the ratchet step $\Delta E$ from the ideal value moves this critical meeting point to lower voltages and therefore decreases the voltage that can be obtained. Decreasing the ratchet step, on the other hand, decreases the available current because the onset of carrier multiplication is moved upward, reducing the number of carriers that are multiplied.

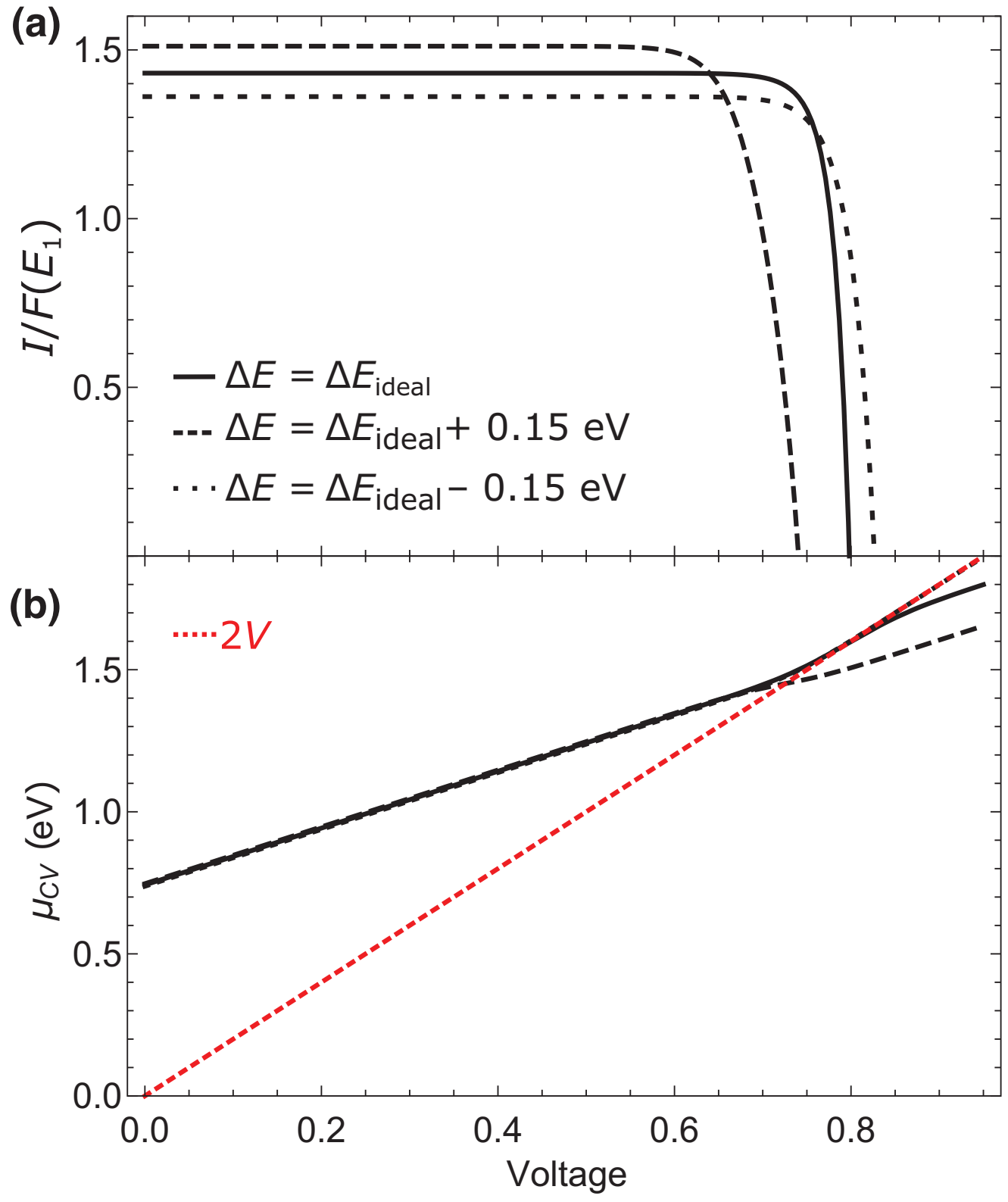

FIG. 7. (a) The $I$-$V$ curves (normalized to the current expected from a cell with a band gap of $E_1$) and (b) the QFLS of the high-energy transition as a function of voltage for a CM solar cell. The curves are for the ideal configuration (solid line) and a configuration with a higher (dashed) or lower (dotted) ratchet step; the fundamental band gap $E_1 = 1.05$ eV (corresponding to the ideal configuration for $w = 1$ mm) is held constant. The red line in (b) indicates double the applied voltage (i.e., the point of no free-energy gain on carrier multiplication).

Interpreting carrier multiplication as the voltage-dependent current source in the equivalent-circuit picture, we can conclude that this current source delivers a current that corresponds to approximately twice the photon flux above the higher band gap. Yet, if the applied voltage becomes too high and the free-energy balance of the impact ionization becomes negative, Auger recombination dominates instead, decreasing the output current.

The $I$-$V$ curve of the AA solar cell Fig. 8 shows a similar behavior. Now the voltage-dependent current source represents the low-energy transition and it delivers a current that corresponds to half the incoming photon flux in its absorption range. For an external short-circuit condition, a QFLS is sustained between the IB and the VB. The Auger process works to increase the overall current, with no change in the QFLS until the applied voltage becomes large enough to enable strong impact ionization. This impact-ionization process maintains the QFLS between the IB and the VB at more than twice the voltage applied across the diode (VB-to-CB transition) as it counteracts the Coulomb-mediated up-conversion of carriers.

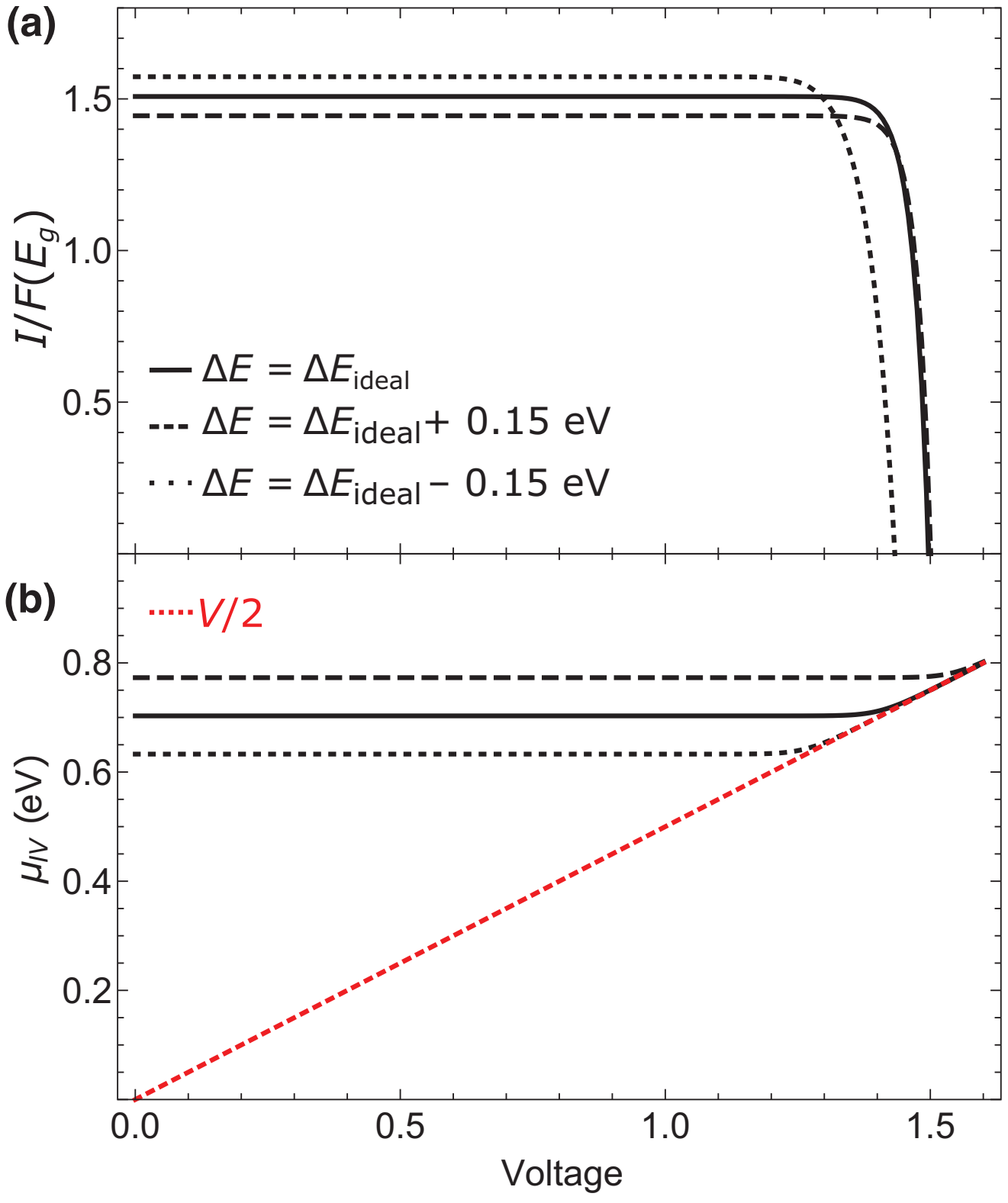

FIG. 8. (a) The $I$-$V$ curves (normalized to the current expected from a cell with a band gap of $E_g$) and (b) the QFLS of the low-energy transition as a function of voltage. The curves are for the ideal configuration (solid line) and a configuration with a higher (dashed) or lower (dotted) ratchet step; the band gap $E_g = 1.81$ (corresponding to the ideal configuration for $w = 1$ mm) is held constant. The red line indicates half the applied voltage (i.e., the line of no free-energy gain on Auger-assisted up-conversion).

For the ideal configuration, the operating point of the cell matches well with the point where the Gibbs free energy created in the Auger process becomes negligible. If the ratchet step is increased beyond the ideal value, the cell loses some of its current, while only marginally increasing voltage. If it is decreased, the cell loses voltage without increasing the current enough to compensate for the voltage loss.

These results illustrate the nature of AA and CM solar cells pertaining to the QFLSs in the device and the ratchet steps. One important aspect that we wish to emphasize is that any device with Auger-type interaction to multiply or up-convert carriers must have a QFLS at short circuit. This should be observable in luminescence experiments, even at short circuit, with appreciable luminescence from the high-energy states in a CM solar cell and appreciable luminescence from the low-energy states in an AA solar cell. At open circuit under solar illumination, luminescence should occur from both thresholds if the material is applicable for parallel solar cells. Such measurements can be performed without contacting the device and can be used for efficient screening of potential materials and devices.

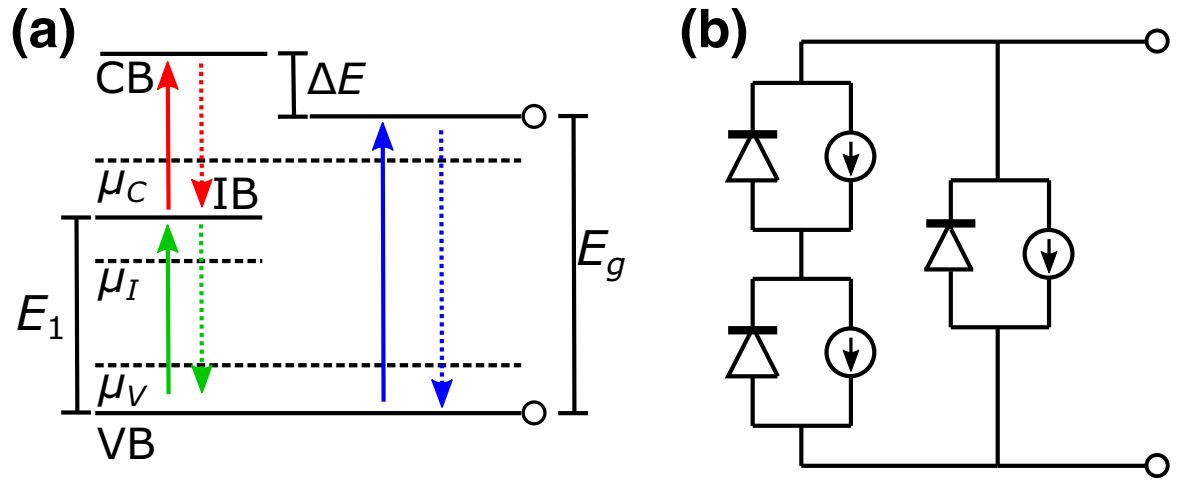

FIG. 9. (a) Band alignment and (b) equivalent circuit of an IBSC.

### C. Intermediate-band solar cell

Despite considerable experimental efforts to realize the IBSC concept [39], the voltage losses of the IBSC compared with the corresponding single-junction device are usually very high and the concept of voltage recovery, with a voltage higher than the lowest absorption threshold, has not yet been proven in experiments [40]. Furthermore, the three-color-luminescence test proposed in Ref. [41] to prove the existence of two separate QFLSs has been passed only at low temperature [42].

The beneficial impact of a ratchet step in IBSCs was discussed in a few publications [10,43,44]. In Sec. II we showed that the requirement for a ratchet step to maximize the matched voltage is the consequence of étendue mismatch between absorption and emission and also possible nonradiative processes. Equation (9) provides the ratchet step required to match the aggregated QFLSs of the individual transitions in the parallel subunits at open circuit. Yet, in an IBSC it is the operating voltage of the series-connected IB transitions, connected in parallel to the VB-to-CB transitions, that needs to be optimized (see equivalent-circuit diagram in Fig. 9). The trade-off involved here leads to numerical values for the ratchet that differ from the value obtained from matching open-circuit voltages.

We consider the asymmetric, sequential-absorption IBSC, with three absorption thresholds, of which the two lower ones involve the IB. Because of the ratchet step, the energies of the two lower absorption thresholds do not have to add up to the VB-CB band gap, which allows freedom to match the internal voltages. Balancing generation and recombination rates for each band as shown in Eqs. (10) and (11), we obtain the open-circuit voltage as a function of the generation rates, saturation currents, and

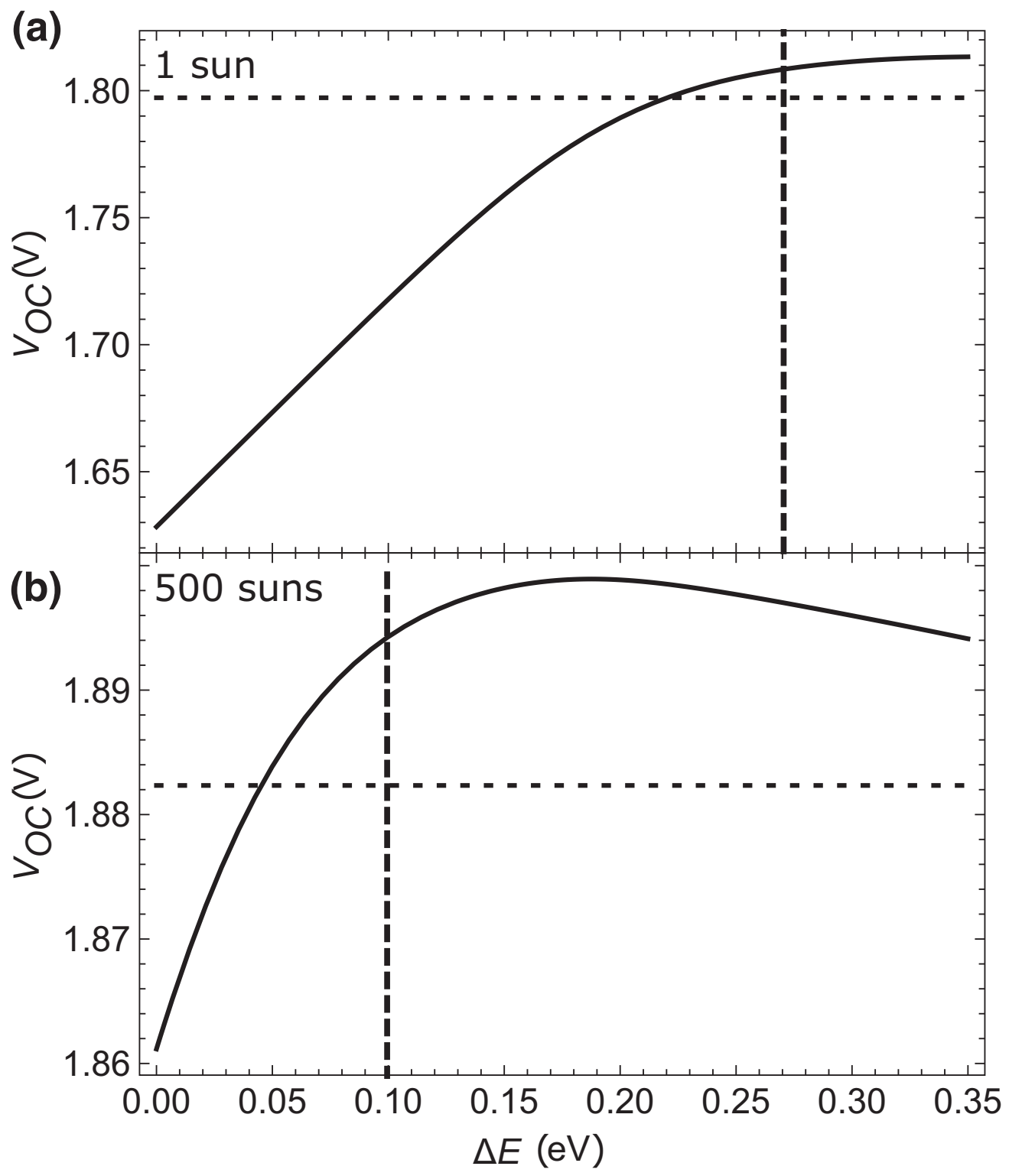

FIG. 10. $V_{\rm OC}$ of the IBSC with the ideal band-gap configuration for (a) 1-sun ($E_g = 2.09$ and $E_1 = 1.42$) and 500-sun ($E_g = 2.01$ and $E_1 = 1.31$) blackbody illumination with $T_S = 6000$ K. The dotted baseline corresponds to $V_{\rm OC}$ of an ideal single-junction solar cell with the same VB-CB band gap. The vertical dashed line indicates the ratchet step that promises the highest efficiency.

QFLSs:

$$V_{\rm OC}^{\rm asym} = \frac{kT_c}{q}\left[\ln f + \ln\left(\frac{G_{IC} - G_{VI} + R^0_{IV} f}{R^0_{CI}}\right)\right], \quad (30)$$

where $f$ is the solution of a quadratic equation given by

$$f = \left(\frac{-b + \sqrt{b^2 + 4ac}}{2a}\right), \quad (31)$$

with

$$a = \frac{R^0_{CV} R^0_{IV}}{R^0_{CI}}, \quad (32)$$

$$b = \frac{G_{IC} - G_{VI}}{R^0_{CI}} R^0_{CV} + R^0_{IV}, \quad (33)$$

and

$$c = G_{VC} + G_{VI}. \quad (34)$$

This equation for the open-circuit voltage applies to any material system where the recombination current of each transition can be approximated with ideality factor 1 and is valid for any ratchet step. In Fig. 10 we nonetheless assume ideal transitions so as to compare the open-circuit voltage of an ideal single-junction solar cell with the open-circuit voltage of an ideal IBSC with the same VB-CB band gap. Figure 10 also shows how the ratchet step at ideal efficiency (shown by the dashed vertical lines) differs from the ratchet step that gives the greatest open-circuit voltage.

The ideal IBSC with the optimal ratchet step, indicated by the dashed line, shows a higher open-circuit voltage than the ideal single-junction solar cell with the same CB-IB separation. Also, the ratchet step needed to recover the open-circuit voltage of the single junction decreases considerably with concentration, as predicted in Eq. (9).

## IV. CONCLUSIONS

In this work, we categorize advanced-concept solar cells into three categories: in series, via luminescence, and in parallel. Étendue mismatch and nonradiative recombination necessitate a degree of freedom in the combination of absorption thresholds that can be provided with a ratchet relaxation step. At 1 sun and for ideal systems this ratchet step is on the order of 270 meV for all the different technologies. The optimal ratchet step generally depends logarithmically on the ratio of the luminescence-extraction efficiencies of the different transitions, just as the sum of the QFLSs does. We explain how the optimal value for this ratchet step depends on device properties and nonidealities for photonic UC and DC solar cells, Auger-type solar cells with CM or internal carrier up-conversion (AA), and sequential-absorption IBSCs and provide illustrative example results for idealized conditions.

We also calculate $I$-$V$ curves for the CM and AA solar cells under the condition of finite Auger interaction rates. While the two concepts are completely symmetric to each other in the regime of infinite interaction rates and therefore yield the same limiting efficiency, finite Auger interaction rates break the symmetry between the two concepts and we find that the CM solar cell is more robust to a slowdown of Auger interaction rates. The ideal ratchet step for the AA solar cell is largely unaffected by the Auger interaction rate, while it decreases sharply with slower interaction in the case of the CM solar cell. Plotting the QFLS of the noncontacted transition against the applied voltage of the device reveals a large QFLS at short circuit, which should lead to appreciable luminescence even at short circuit in a functional Auger-type solar cell.

IBSCs require voltage matching through ratchets for their optimal functioning, especially in the presence of nonidealities. We show that there is a small difference between voltage matching at open circuit and voltage matching at the operating point of IBSCs, the latter being the important consideration.

Finally, controlling the QFLSs is important for efficient operation of any member of the parallel class of solar cells. The existence of multiple QFLSs in a candidate material should be confirmed by luminescence experiments.

## ACKNOWLEDGMENTS

The authors acknowledge valuable discussions with M.J.Y. Tayebjee, I. Perez-Wurfl, S. Bremner, T.W. Schmidt, and U. Römer.